\documentclass[notitlepage,twocolumn,aip,hidelinks,10pt,byrevtex,secnumarabic,nofootinbib,longbibliography,a4paper,showkeys,preprintnumbers]{revtex4-2}
\usepackage{graphicx}
\usepackage{amsmath}
\usepackage{amssymb}
\usepackage{float}
\usepackage{censor}
\usepackage{moreverb}
\usepackage{url}
\usepackage{tabularx,array}
\usepackage{xr-hyper}
\usepackage{xcolor}
\usepackage{colortbl}
\usepackage[draft,commandnameprefix=always]{changes}
\usepackage{tikz}
\usepackage{pgfplots}

\usetikzlibrary{shapes,arrows,arrows.meta,positioning,calc,external,patterns}
\usepgfplotslibrary{fillbetween}
\pgfplotsset{compat=1.11}

\usepackage[pdftex,
pdfauthor={Jose-Maria Martin-Olalla},
pdftitle={Proof of the Nernst theorem},
            pdfsubject={Foundations of Thermodynamics},
            pdfkeywords={second law of thermodynamics; third law of thermodynamics; specific heat; statistical mechanics; entropy; temperature; Carnot engine; foundations of thermodynamics; Carnot theorem; absolute zero},
            pdfproducer={Latex with hyperref, or other system},
            pdfcreator={pdflatex, or other tool}]{hyperref}

                        \newcommand\rorlink[2]{\href{https://ror.org/#2}{#1\,\includegraphics[scale=0.1]{ror-icon-rgb}}
}
\usepackage{orcidlink}
\providecommand{\orcidlinki}[2]{\href{https://orcid.org/#2}{#1}\orcidlink{#2}}
            \usepackage{breakurl}
\usepackage{siunitx}
\usepackage{mathtools}
\usepackage[utf8]{inputenc}
\usepackage[english]{babel}
\usepackage{lineno}
\usepackage{array}

\newcommand{\uQ}{\ensuremath{Q}}

\newcommand{\Qrev}{\ensuremath{\mathsf{\uQ}}}

\AtBeginDocument{\usepackage{booktabs}}

\graphicspath{{./figures/}}
\makeatletter\AtBeginDocument{\let\@elt\relax}\makeatother

\newcommand{\uaddress}{\rorlink{Universidad de Sevilla}{03yxnpp24}, Facultad de Física, Departamento de Física de la Materia Condensada,  ES41012 Sevilla, Spain}

\begin{document}
\author{\orcidlinki{José-María Martín-Olalla}{0000-0002-3750-9113}}
\affiliation{\uaddress}
\email{olalla@us.es}

\title{Proof of the Nernst heat theorem}
\received[Submitted ]{19 March 2024}
\revised{19 February 2025}
\accepted{01 June 2025}
\published{13 June 2025}
\preprint{\textcolor{blue}{This is the Author's accepted manuscript. Traducción a la lengua castellana en página 5 y ss.  \copyright The Author, 2025}}
\preprint{\textcolor{blue}{The Version of Record is published by \emph{The European Physical Journal Plus} \textbf{140} 528 \doi{10.1140/epjp/s13360-025-06503-w}}}
\begin{abstract}

  The Nernst heat theorem is proven from purely thermodynamic arguments connected with the second law of thermodynamics. The proof stipulates that $T=0$ is formalized by a Carnot thermometer, and is independent of the vanishing of the specific heats, or the unattainability of the zeroth isotherm.

  With this proof, the second law of thermodynamics would extend its applicability and the third postulate of thermodynamics would be narrowed to the fact that the entropy of a finite-density, chemically homogeneous body must not be negative.


\end{abstract}


\maketitle

\section{Introduction}
\label{sec:introduction}

In classical thermodynamics, the Nernst theorem or Nernst heat theorem states that \emph{the isothermal change of entropy $(\Delta S)_T$ associated with a process between two states of a system in internal equilibrium vanishes as the temperature vanishes}.\cite{epstein-1937,pippard-1957,hatsopoulos-1965,kestin-1968ii} It is one of the two general properties of matter in the vicinity of the zeroth isotherm.  Formally, the theorem sets the vanishing of $(\partial S/\partial X)_t=(\partial^2A/\partial T\partial X)$ when $T\to0^+$, where $X$ is a suitable mechanical parameter such as the pressure, volume or magnetic field, and $A(T,X)$ is a suitable thermodynamic potential like the free energy or the free enthalpy. Note that the formulation of the theorem excludes mixing processes, and the variation of entropy associated with them.

Nernst presented his theorem in 1905 after empirical evidence related to chemical equilibrium and to the expansion coefficients at very low temperatures.\cite{nernst-1906} Soon afterwards, he collected evidence on the vanishing of the specific heats as the temperature vanishes,\cite{nernst-1912} a second general property of matter at very low temperatures---related to $(\partial^2A/\partial T^2)$---, further completed by Simon.\cite{Simon1937} Eventually Nernst proposed the unattainability of the zeroth isotherm as a summary of his evidences and presented a proof by contradiction.\cite{nernst-1912} Nernst maintained that since a Carnot engine operating at the zeroth isotherm would negate the second law, the unattainability of the zeroth isotherm is then deduced.

Einstein refuted the proof noting that at $T=0$ every irreversibility, no matter how small, would throw the system away from the zero isotherm, paving the way for the third law of thermodynamics.\cite[p.~293]{Solvay-1913} Later, Epstein refined the rebuttal saying that at $T=0$ the isothermal process becomes also adiabatic and therefore no practical procedure would be able to accomplish this process. Hence the engine would not operate and could not possibly challenge the second law.\cite{epstein-1937,kestin-1968ii,Kox2006,Boas1960} No other attempt to probe the theorem is known.

The isothermal change of entropy plays a crucial role in reversible Carnot engines, which on their own play a crucial role in the second law of thermodynamics. A reversible Carnot engine extracts an equivalent heat $\mathsf{Q_h}/T_h$(sans-serif letters will be used to identify exchanges in reversible processes) from a heater at temperature $T_h$ and pours it into the cooler. For that to occur the substance undergoing the cycle must be able to sustain an isothermal change of entropy $(\Delta S)_T=\mathsf{Q_h}/T_h$ at the temperature of the heater and at the temperature of the cooler $T_c$, both conducted by an isothermal change of the mechanical parameter $X$. The engine produces work $\mathsf{W}=(\Delta S)_T\times(T_h-T_c)$.

This work presents a proof of the Nernst theorem purely based on thermodynamic arguments connected with the fact that the Carnot thermometer must be able to operate at $T=0$. The proof does not require the vanishing of the specific heat. It is not a proof by contradiction, like the proof presented by Nernst, but of consistency of the second law.

\section{Background}
\label{sec:background}

Planck's statement of the second law of thermodynamics reads  \emph{it is impossible to construct an engine which will work in a complete cycle, and produce no effect except the raising of a weight and the cooling of a heat-reservoir}\cite{planck-1897}. It formalizes an observation by \citet{carnot-1824}: ``the production of heat alone is not sufficient to give birth to the impelling power; it is necessary that there should also be cold; without it, the heat would be useless.'' Consequently, an engine cannot operate with a solo reservoir (unary engines), instead,  it must heat, at least, a second, distinct reservoir, thus setting a binary engine.

The statement can be formalized as:
\begin{subequations}
   \label{eq:2}  
\begin{align}
   \mathrm{P}\Longrightarrow&(W>0\Rightarrow Q_c<0),   \label{eq:2a}\\
   \neg\mathrm{P}\Longleftarrow&(W>0 \land Q_c\geq0).   \label{eq:2b}
 \end{align}
\end{subequations}
It reads: Planck's statement $\mathrm{P}$ implies that if a weight is lifted ($W>0$, antecedent), then a reservoir must be heated ($Q_c<0$, consequent); if a weight is lifted and ($\land$) no reservoir is heated, then Planck's statement is negated ($\neg\mathrm{P}$). This logical structure provides a basis for proving Carnot's theorem and Clausius's theorem.

The very notion of temperature is alien to the conservation of energy (in this context $W=Q_h+Q_c$, where $Q_h>0$ is the heat supplied by the heat-reservoir which is cooled) and to Planck's statement\cite{planck-1897}. Instead, Carnot's theorem shows that in reversible binary engines (Carnot engines) the ratio of the exchanged heats is independent of the engine itself, and can be expressed as the ratio of a quantity that depends only on the two reservoirs:
\begin{equation}
  \label{eq:6}
  \frac{T_c}{T_h}=-\frac{\mathsf{Q}_c}{\mathsf{Q}_h},
\end{equation}
 Because the heat exchanges point in opposition, the minus sign in equation~(\ref{eq:6}) ensures that the ratio $T_c/T_h$ is positive. With $W>0$ energy conservation mandates $\mathsf{Q}_h>-\mathsf{Q}_c$, therefore $T_h>T_c$, which aligns with the intuitive idea of hotness and coldness. Equation~(\ref{eq:6}) formally introduces the concept of temperature and  provides a universal thermometer: the Carnot thermometer.\cite{CPS-1848}

\section{The proof}
\label{sec:proof}

In the context presented in section~\ref{sec:background} if $T=0$ is assigned to a finite-density substance, then a reversible Carnot engine operating with a zero temperature ``cooler'' and a heater must be conceptually admitted. Otherwise, the assignment $T=0$ would be external to the framework of the second law that brings Planck's statement. 

This reversible Carnot engine must exchange no heat with the cold reservoir since, from equation~(\ref{eq:6}), if $T_c=0$, then $\Qrev_c=0$.  It then follows that the engine would negate Planck's statement unless no heat were exchanged with the heater: $\Qrev_h=0$, which would result in $W=0$.  Hence, since $T_h\neq0$, we have $\mathsf{Q}_h/T_h=0$ . Therefore, from equation~(\ref{eq:6}) we arrive at $\Qrev_c/T_c=0$, for $T_c=0$.

Because of the universality of Carnot theorem, this result must apply irrespective of the body of finite density that undergoes the cycle and, given a body, irrespective of the specific mechanical or chemical configuration under which the engine operates. Because $\Qrev/T$ exchanged by the reversible engine matches with $(\Delta S)_T$ in the substance undergoing the cycle, it then follows that every $(\Delta S)_T$ at $T=0$  must be zero for every body of finite density, irrespective of the change in the configuration of the body. Continuity then requires that $(\Delta S)_T$ vanishes as the temperature vanishes, which is the Nernst theorem.

Alternatively, a literal reading of Planck's statement may consider that the other effect might not be energetic of necessity. As an example, a reversible Carnot engine lifting a weight and operating at $T=0$ ---where $\Qrev_c=0$---  would not negate the statement because $\Qrev_h/T_h$ would be poured into the $T=0$ reservoir, being that the required additional effect. This reading is  implying that the isothermal change of entropy at $T=0$ can be non-zero. Consequently, it fails to align with the empirical observations underpinning the Nernst theorem, and is therefore discarded.

 \section{Corollaries}
\label{sec:consecuences-proof}

Two well-known corollaries of the Nernst theorem would now follow from the second law. 

First, $S(T=0,X)$ is unique, irrespective of $X$, should $\lim_{T\to0}S(T,X)$ exist. It would be the only circumstance in which the second law points toward absolute $S$, instead of $\Delta S$, albeit incompletely.

Second, irrespective of $\lim_{T\to0}S(T,X)$, it is the case that $S(T,X)>S(0,X)$ for any non-zero $T$. Since the entropy must not decrease adiabatically, no adiabatic cooling can bring the temperature of a finite density system arbitrarily close to $T=0$. This is the statement of unattainability of the zeroth isotherm.

\section{Remark on the absolute value of entropy}
\label{sec:consequences}

The most comprehensive summary of Nernst's empirical observations is due to Planck, who presented the following statement: \emph{as the temperature diminishes indefinitely, the entropy of a chemical homogeneous body of finite density approaches indefinitely near to the value zero}\cite[Part~\textsc{IV}, Chapter~\textsc{VI}]{planck-1911}\cite{hatsopoulos-1965,kestin-1968ii} This concerns absolute entropy, not entropy change.

While the statement is presented as one law (the third law of thermodynamics), it encompasses the two distinct, general properties of matter in the vicinity of the zeroth isotherm: (1) the entropy is unique (as deduced from the vanishing of $(\Delta S)_T$), and (2) the entropy has a floor, which can be set at zero, as deduced from the fast vanishing of the specific heats $c_x$ as the temperature vanishes in view that the integrand in $S=\int c_x \mathrm{d}T/T$ remains bounded.

The fast vanishing of the specific heats is unrrelated to the arguments presented in the proof of the Nernst theorem. Therefore it is the one and only general property of finite density systems that remains independent from the second law. It can be summarized by the following statement: \emph{the entropy of a finite-density, chemically homogeneous body is not negative}.

\section{Discussion}
\label{sec:remarks-proof-nernst}

The burden of the proof lies in the meaning of the assignment $T=0$. Earlier in the 18th century, the absolute zero was empirically conceived as a scenario where the volume or the pressure of a gas vanished. In the argument between Nernst and Einstein, the absolute zero meant $p=0$\cite[page~294]{Solvay-1913}.  From a classical kinematic point of view, $T=0$ is conceived as the state where motion, kinetic energy, vanishes. In some axiomatic presentations of the thermodynamics, $T=0$ is conceived as a scenario where $(\partial U/\partial S)_X$ vanishes ($U$ is the energy of the system).\cite{callen-85,Beretta2015e}

Yet, in the classical framework derived from the second law, equation~(\ref{eq:6}) provides a natural zero for the temperature, associated with $\mathsf{Q}_c=0$. This is a formal assignment of general validity, beyond the empirical $p=0$ or $V=0$, and prior to the definition of the entropy in the classical presentation of thermodynamics. The above proof commits to this association and makes a consistent use of that. It aligns with the proof of the theorems that are deduced from the second law of thermodynamics, such as Carnot theorem or Clausius theorem, linking general properties of cycles with general properties of substances undergoing the cycle. 

In contrast, under Einstein's rebuttal of Nernst's proof ---a Carnot engine operating at $T=0$ can not be constructed---  it follows that $T=0$ can not be determined by equation~(\ref{eq:6}), which is paramount for defining the temperature, but externally, ie empirically.

\begin{figure}
  \centering
  \begin{tikzpicture}
\begin{axis}[name=plot1,width=7.5cm,height=6.0cm,
      samples=100,
      ytick=\empty,
      axis on top,
      xtick={0},
      xticklabels={$T=0$},
      ytick={0.05},
      yticklabels={$\mathrm{E}_0$},
      domain=1:3,
      axis x line=bottom,
      axis y line=left,
      ymin=0,
      ymax=1,
      xmin=0,
      xmax=1,
      xlabel=$T$,
      ylabel=$S$,
      every axis x label/.style={at={(ticklabel* cs:1.05)}},
      every axis y label/.style={at={(ticklabel* cs:1.05)}}]

      \coordinate (E0) at (0,0.05);
      \coordinate (E1) at (0.7,0.8);
      \coordinate (E2) at (0.2,0.15);
      \coordinate (E3) at (0.4,0.6);
      \coordinate (E4) at (0.4,.95);
      \coordinate (E5) at (0,.95);
      \path [name path=axisy] (axis cs:0,0) -- (axis cs:0,0.95);
      \path [name path=axisx] (axis cs:0,0) -- (axis cs:.95,0);
      \addplot [gray!50!white,name path=A,domain=0:0.45] {2*x+0.05};
      \addplot [gray!50!white,name path=B,domain=0:0.95] {.2*x+0.05};
      \addplot[gray,semitransparent] fill between [of=A and axisy];
      \addplot[gray,semitransparent] fill between [of=B and axisx];
      \draw [thick,latex-latex] (E0) .. controls  (E2) and (E3) .. (E1);
      \fill (E1) circle (1pt);
      \node [right] at (E1) {$\mathrm{E}_1$};
    \end{axis}
  \end{tikzpicture}
  \caption{Any reversible Carnot engine operating at $T=0$ must consist of a round trip from a given equilibrium state $\mathrm{E}_1$ to $\mathrm{E}_0$ at $T=0$. Following the vanishing of $c_x$, the entropy at $T=0$ is placed at a finite value of $S$.}
  \label{fig:nernst}
\end{figure}
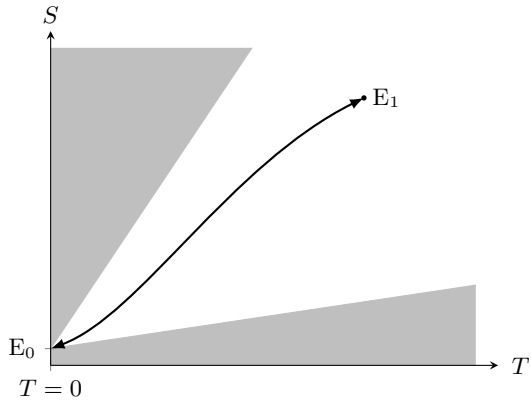

In summary, $T=0$ would be assigned to a finite system only if the reversible Carnot engine that attempts to assess the temperature of the system consists of a round trip connecting a given state $\mathrm{E}_1$ of the working fluid with the state $\mathrm{E}_0$, see figure~\ref{fig:nernst}.  The loop, purely conceptual, will enclose no area, no work will be produced, and no entropy will be exchanged.\footnote{The Carnot engine is usually shown as a rectangle in a $ST$ chart. Actually a Carnot engine ---an engine operating with two distinct heat sources--- requires two isothermal processes and two parallel paths shifted by an entropy $\Lambda$. In the limit $\Lambda\to0$ the paths coalesce in one round trip path as shown in Figure~\ref{fig:nernst}.\cite{Martin-Olalla2003b}.  Note that, conceptually, the engine would keep operating with $T_c=0$. It would only cease to deliver any work.}

The interplay between  Planck's statement of the second law and the Nernst theorem (N) is also worthy of discussion.

The Nernst theorem can be described by quantities related to reversible Carnot engine. First, the limit $T\to0$ is the limit $\mathsf{Q}_c\to0^-$, as per Eq.~(\ref{eq:6}). Second, the observation $(\Delta S)_T\to0$ is translated into $\mathsf{Q}_h\to0^+$ and into $\mathsf{W}\to0^+$. Therefore the Nernst theorem expresses the following relations:
\begin{subequations}
   \label{eq:4}
 \begin{align} 
\mathrm{N}\Longrightarrow& \big((\mathsf{Q}_c\to0^-)\Rightarrow(\mathsf{W}\to0^+) \big),   \label{eq:4a}\\
\neg\mathrm{N}\Longleftarrow&\big( (\mathsf{Q}_c\to0^-)\land \mathsf{W}>0\big).   \label{eq:4b}
 \end{align}  
\end{subequations}

Comparing with the right material implication in~(\ref{eq:2a}), the right material implication in~(\ref{eq:4a}) is saying: the consequent ($\mathsf{Q}_c$) vanishes only if the antecedent ($\mathsf{W}$) vanishes as well.

On the other hand, the left material implications in~(\ref{eq:2b}) and in~(\ref{eq:4b}) differ slightly. In (\ref{eq:2b}), $\mathsf{Q}_c$ is seen as a categorical variable: if $\mathsf{Q}_c=0$ and $\mathsf{W}>0$, then the statement is negated; if $\mathsf{Q}_c$ is negative, then the statement survives, no matter how close to zero $\mathsf{Q}_c$ might be. Einstein's and Epstein's refutations of Nernst's proof is in line with this view: it is impossible to construct an engine operating at $T=0$ but, for any $T>0$, no matter how close to zero it might be, the engine could be constructed as the categorical argument presented by them no longer sustains; nonetheless, Planck's statement would sustain as long as $\mathsf{Q}_c<0$.\footnote{Since the isothermal change of entropy at the cooler is negative, the following categorical argument is also possible. For $T_c=0$ the isothermal process is also adiabatic and since the entropy must not decrease adiabatically, the engine could not be constructed. At any $T_c\neq0$ the argument no longer applies because the isotherm is now diathermal and the entropy of the substance can indeed decrease isothermally, irrespective of how close to zero $T_c$ might be.}

 In contrast, proposition~(\ref{eq:4b}) brings $\mathsf{Q}_c$ as a continuous variable. In this line, the Nernst theorem acknowledges that the other energetic effect $\mathsf{Q}_c$ that appears in the conversion of heat into work is not categorical but, like any other mandatory effect, a mensurable quantity, larger than a value prescribed by nature through $\mathsf{Q}_h/T_h$ (the entropy carried by the engine) and the thermophysical properties of the substance undergoing the cycle.\cite{Martin-Olalla2003b}

%

\appendix

\section{History of the manuscript}
\label{sec:history-manuscript}

This proof originated in the writing of a textbook on thermodynamics that started in the Fall of 2023. During the Christmas holiday of 2023 I was jotting down the consequences of the Planck statement of the second law, including the concept of temperature. I realized the importance of the natural zero of temperature in regard to the Nernst theorem and outlined the proof immediately.

The manuscript was posted in 8 January 2024 \doi{10.48550/arXiv.2401.04069} and submitted to top tier journals.

The proof was first presented on 6 November 2024 when addressing the consequences of the Planck statement of the second law to the sophomore students in the course on thermodynamics I regularly teach.

\end{document}


\author{\orcidlinki{José-María Martín-Olalla}{0000-0002-3750-9113}}
\affiliation{\uaddress}
\email{olalla@us.es}

\title{Demostración del teorema de Nernst}
\received[Enviado ]{19 de marzo del 2024}
\revised[Revisado ]{19 de febrero del 2025}
\accepted[Aceptado ]{01 de junio del 2025}
\published[Publicado ]{13 de junio del 2025}
\preprint{\textcolor{blue}{Esta es la traducción a la lengua castellana. \copyright The Author, 2025}}
\preprint{\textcolor{blue}{The Version of Record is published by \emph{The European Physical Journal Plus} \textbf{140} 528 \doi{10.1140/epjp/s13360-025-06503-w}}}
\begin{abstract}
Se demuestra el teorema del calor de Nernst a partir de argumento exclusivamente termodinámics relacionados con el segundo principio de la termodinámica. La demostración estipula que $T=0$ tiene que ser establecido por un termómetro de Carnot, y es independiente de la anulación de los calores específicos y de la inaccesibilidad de la isoterma cero.

Con esta demostración, el segundo principio de la termodinámica extendería su rango de aplicabilidad y el tercer postulado quedaría reducido al hecho de que la entropía de una sustancia químicamente homogénea y de densidad finita no puede ser negativa.
\end{abstract}


\maketitle

\section{Introduction}
\label{sec:introduction}

En termodinámica clásica, el teorema de Nernst o teorema del calor de Nernst establece que \emph{el cambio isotermo de entropiá $\Delta S_t$ asociado a un proceso entre dos estados de un sistema en equilibrio interno tiende a cero cuando la temperatura tiende a cero}.\cite{epstein-1937,pippard-1957,hatsopoulos-1965,kestin-1968ii} Es la primera de las dos propiedades generales de la materia en las cercanías de la isoterma cero. Formalmente, el teorema implica la anulación de  $(\partial S/\partial X)_t=(\partial^2A/\partial T\partial X)$ cuando $T\to0^+$, donde $X$ es un parámetro mecánico adecuado como la presión, el volumen, o el campo magnético; y  $A(T,X)$ es un potencial termodinámico adecuado, como la energía libre o la entalpía libre. Observe que la formulación del teorema excluye procesos de mezcla y la variación de entropía asociada a ellos.

Nernst presentó su teorema en 1905 tras obtener indicios relacionados con el equilibrio químico y con los coeficientes de expansion a muy baja temperatura.\cite{nernst-1906} Poco después, obtuvo indicios de la anulación de los calores específicos cuando la temperatura tiende a cero,\cite{nernst-1912} una segunda propiedad general de la materia a muy baja temperatura---relacionada con $(\partial^2A/\partial T^2)$---, que fueron completados por Simon.\cite{Simon1937} Como conclusión Nernst propuso la inaccesibilidad de la isoterma cero como resumen de sus indicios y propuso una demostración por contradicción de su teorema.\cite{nernst-1912} Nernst sostuvo que dado que una máquina de Carnot operando en la isoterma cero violaría el segundo principio de la termodinámica, se deduce entonces la inaccesibilidad de la isoterma cero.

Einstein refutó la demostración haciendo ver que en $T=0$ cualquier irreversibilidad o imperfección, por pequeña que fuera, alejaría al sistema de la isoterma cero y despejó el camino para un tercer principio de la termodinámica.\cite[p.~293]{Solvay-1913} Más tarde, Epstein mejoró la refutación señalando que en la isoterma cero los procesos isotermos son también adiabáticos y, por tanto, no hay forma práctica de conducir el ciclo de Carnot. Si la máquina no puede operar, entonces no puede cuestionar la validez del segundo principio de la termodinámica.\cite{epstein-1937,kestin-1968ii,Kox2006,Boas1960} No se conoce ningún otro intento de demostrar el teorema.

El cambio isotermo de entropía desempeña un papel crucial en las máquinas de Carnot reversibles, que, a su vez, desempeñan un papel crucial en el segundo principio de la termodinámica. Una máquina reversible de Carnot extrae un equivalente del calor $\mathsf{Q_h}/T_h$ (las letras en palo seco representan intercambios reversibles) de la caldera a temperatura $T_h$ y lo deposita en el refrigerador. Para que esto ocurra, la sustancia que realiza el ciclo termodinámica debe ser capaz de soportar un intercambio isotermo de entropía $(\Delta S)_t=\mathsf{Q_h}/T_h$ tanto a la temperatura de la caldera como a la temperatura del refrigerador $T_c$, ambos cambios son dictados por el cambio isotermo del parámetro mecánico $X$. La máquina produce un trabajo igual a: $\mathsf{W}=(\Delta S)_T\times(T_h-T_c)$.

Este estudio presenta una demostración del teorema de Nernst basada solamente en argumentos termodinámicos relacionados con el hecho de que el termómetro de Carnot debe ser capaz de operar en $T=0$. La demostración no requiere la anulación de los calores específicos. No es una demostración por contradicción, como la propuesta por Nernst, sino de consistencia del segundo principio.

\section{Marco teórico}
\label{sec:background}

El postulado de Planck del segundo principo de la termodinámica dice: \emph{es imposible construir una máquina que trabaje de forma cíclica y no produzca otros efectos excepto la elevación de una carga y el enfriamiento de un repositorio de calor}\cite{planck-1897}. El postulado formaliza una observación de \citet{carnot-1824}: "<la producción de calor por sí sola no es suficiente para dar lugar a la fuerza motriz; es necesario también que haya frío; sinél, el calor sería inútil."> Por tanto, una máquina no puede operar con un único reporsitorio de calor (máquina unaria), por contra, debe calentar, al menos, a un segundo y distinto repositorio de calor, conformando así una máquina binaria.

El postulado puede formalizarse como:
\begin{subequations}
   \label{eq:2}  
\begin{align}
   \mathrm{P}\Longrightarrow&(W>0\Rightarrow Q_c<0),   \label{eq:2a}\\
   \neg\mathrm{P}\Longleftarrow&(W>0 \land Q_c\geq0).   \label{eq:2b}
 \end{align}
\end{subequations}
Que se lee así: el postulado de Planck $\mathrm{P}$ implica que si una carga se eleva ($W>0$, antecedente), entonces un repositorio ha de ser calentado ($Q_c<0$, consequente); si una carga se eleva y and ($\land$) ningún repositorio se calienta, entonces el postulado se niega ($\neg\mathrm{P}$). Esta estructura lógica proporciona las bases para demostrar el teorema de Carnot y el teorema de Clausius.

La propia noción de temperatura es ajena a la conservación de la energía (en este contexto $W=Q_h+Q_c$, donde $Q_h>0$ es el calor suministrado por el repositorio que se enfría) y el postulado de Planck.\cite{planck-1897} Por cotnra, el teorema de Carnot muestra que en máquinas binarias reversibles (máquinas de Carnot) la razón entre los calores intercambiados es independiente de la máquina, y puede ser expresada como la razón de una cantidad que depende exclusivamente de las propiedades de los dos repositorios:
\begin{equation}
  \label{eq:6}
  \frac{T_c}{T_h}=-\frac{\mathsf{Q}_c}{\mathsf{Q}_h},
\end{equation}
Como los intercambios de calor apuntan en direcciones contrarias, el signo menos de la ecuación~(\ref{eq:6}) asegura que la razón $T_c/T_h$ es positiva. Con $W>0$, la conservación de la energía obliga a que $\mathsf{Q}_h>-\mathsf{Q}_c$, y por tanto $T_h>T_c$, que se corresponde con la idea intuitiva de calor y frío. La ecuación~(\ref{eq:6}) presenta formalmente el concepto de temperatura y proporciona un termómetro universal: el termómetro de Carnot.\cite{CPS-1848}

\section{La demostración}
\label{sec:proof}

En el contexto presentado en la sección~\ref{sec:background} si se asigna $T=0$ a una sustancia de densidad finita, entonces una máquina de Carnot reversible que opere con un ``refrigerador'' a la temperatura cero debe ser admitida conceptualmente. Si no, la asignación $T=0$ sería externa al marco formal del segundo principio que nace tras el postulado de Planck.

Esta máquina de Carnot reversible no puede intercambiar calor con el respositorio frió puesto que, de la ecuación~(\ref{eq:6}), se tiene que si $T_c=0$, entonces $\Qrev_c=0$.  Se deduce entonces que la máquina negaría el postulado de Planck excepto si tampoco se intercambiara calor con la caldera: $\Qrev_h=0$, lo que resultaría en $W=0$.  Como $T_h\neq0$, se tiene que $\mathsf{Q}_h/T_h=0$ . Por tanto, de la ecuación~(\ref{eq:6}) se llega a $\Qrev_c/T_c=0$, para $T_c=0$.

Dada la universalidad del teorema de Carnot, este resultado es válido con independencia de la sustancia de densidad finita que realice el ciclo termodinámico, y para una sustancia dada, es independiente de la configuración mecánica concreta bajo la que opere la máquina. Como el $\Qrev/T$ intercambiado por una máquina coincide con $(\Delta S)_t$ en la sustancia que realiza el ciclo, se deduce entonces que cualquier $(\Delta S)_T$ en $T=0$  debe ser cero para cualquier sustancia de densidad finita, con independencia del cambio en la configuración que ocurra en la sustancia. La continuidad entonces requiere que $(\Delta S)_T$ se anula conforme la temperatura se anula, que es el teorema de Nernst.

Alternativamente, una lectura literal del postulado de Planck podría considerar que el otro efecto al que alude el postulado no tiene que ser necesariamente energético. Por ejemplo, una máquina de Carnot reversible que eleve una carga operando con $T=0$ ---donde $\Qrev_c=0$--- no negaría el postulado porque $\Qrev_h/T_h\neq0$ sería suministrado en el refrigerador, y sería el efecto adicional necesario. Esta lectura implica que el cambio isotermo de entropía en $T=0$ puede ser distinto de cero. Por tanto, no es compatible con las observaciones empíricas relacionadas con el teorema de Nernst, y debe ser descartada.

 \section{Corolarios}
\label{sec:consecuences-proof}

Dos conocidos corolarios de teorema de Nernst se deducirían ahora del postulado de Planck.

Primero, $S(T=0,X)$ es un valor único, independiente de $X$, siempre que exista $\lim_{T\to0}S(T,X)$. Sería la única circunstancia en la que el segundo principio apunta a la entropía absoluta, y no al cambio de entropía, aunque lo haría de forma incompleta.

Segundo, con independencia de $\lim_{T\to0}S(T,X)$, siempre es el caso que  $S(T,X)>S(0,X)$ para cualquier temperatura no nula. Como la entropía no puede disminuir en un proceso adiabático, se deduce que ningún enfriamiento adiabático puede reducir la temperatura de un sistema de densidad finita a un valor arbitrariamente próximo a $T=0$. Este es el enunciado de inaccesibilidad de la isoterma cero.

\section{Apunte sobre el valor absoluto de la entropía}
\label{sec:consequences}

El resumen más completo de las observaciones empíricas de Nernst fue escrito por Planck, mantuvo que \emph{cuando la temperatura disminuye indefinidamente, la entropía de una sustancia químicamente homogénea se aproxima indefinidamente a un valor cercano a cero}\cite[Part~\textsc{IV}, Capítulo~\textsc{VI}]{planck-1911}\cite{hatsopoulos-1965,kestin-1968ii} Este enunciado concierne a la entropía absoluta, y no al cambio de entropía.

Aunque este enunciado se presenta como una única ley (el tercer principio) incluye dos propiedades generales de la materia en la cercanía de la isoterma cero que son distintas:  (1) la entropía es única (como se deduce de la anulación de  $(\Delta S)_t$), y (2) la entropía tiene un suelo, que puede establecerse en cero, como se deduce de la anulación rápida de los calores específicos viendo que el integrando de  $S=\int c_x \mathrm{d}T/T$ permanece finito.

La anulación rápida de los calores específicos no está relacionada con los argumentos presentados para demostrar el teorema de Nernst. Por tanto, es la única propiedad de los sistemas de densidad finita que es independiente del postulado de Planck del segundo principio. Puede sintetizarse en el siguiente postulado: \emph{la entropía de un sistema de densidad finita y químicamente homogéneo no puede ser negativa.}

\section{Discusión}
\label{sec:remarks-proof-nernst}

El fundamento de la demostración reside en el significado de la asignación $T=0$. Ya en el siglo 18º, el cero absoluto fue concebido empíricamente como la situación en la que el volumen o la presión del gas se anulaban. En la discusión entre Nernst y Einstein, el cero absoluto significaba $p=0$\cite[pág~294]{Solvay-1913}.  Desde un punto de vista cinemático, $T=0$ es el estado donde el movimiento, la energía cinética, desaparece. En algunas presentaciones axiomáticas de la termodinámica, $T=0$ se concibe como la situación en la que $(\partial U/\partial S)_X$ se anula ($U$ es la energía del sistema).\cite{callen-85,Beretta2015e}

Sin embargo, en el marco clásico del segundo principio la ecuación~(\ref{eq:6}) proporcional un cero natural para la tempratura, asociado a $\mathsf{Q}_c=0$. Esta asignación formal tiene validez general, más allá de las ecuaciones empíricas $p=0$ o $V=0$, y es anterior a la definición de entropía en la presentación clásica de la termodinámica. La demostración anterior se ajusta a esta idea y hace un uso consistente de ella. Se alinea con las demostraciones de los teoremas que se deducen del segundo principio de la termodinámica, como el teorema de Carnot o el de Clausius, relacionando propiedades generales de los ciclos termodinámicos con propiedades generales de las sustancias que realizan los ciclos. 

En contraste, bajo la refutación de Einstein del la demostración de Nernst ---una máquina de Carnot operando en $T=0$ no puede construirse---  se deduce que $T=0$ no puede ser determinada por la ecuación~(\ref{eq:6}), que es clave para definir el concepto de temperatura, sino que ha de hacerse externamente, es decir empíricamente.

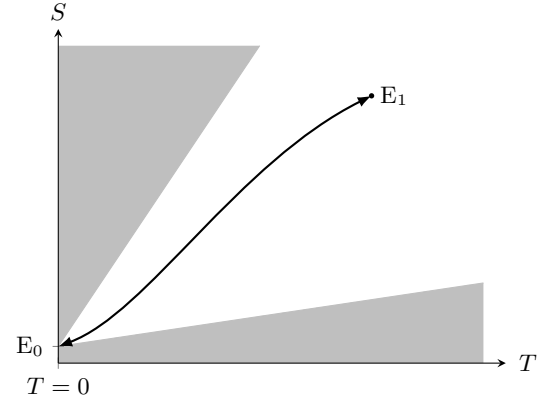
\begin{figure}
  \centering
  \begin{tikzpicture}
\begin{axis}[name=plot1,width=7.5cm,height=6.0cm,
      samples=100,
      ytick=\empty,
      axis on top,
      xtick={0},
      xticklabels={$T=0$},
      ytick={0.05},
      yticklabels={$\mathrm{E}_0$},
      domain=1:3,
      axis x line=bottom,
      axis y line=left,
      ymin=0,
      ymax=1,
      xmin=0,
      xmax=1,
      xlabel=$T$,
      ylabel=$S$,
      every axis x label/.style={at={(ticklabel* cs:1.05)}},
      every axis y label/.style={at={(ticklabel* cs:1.05)}}]

      \coordinate (E0) at (0,0.05);
      \coordinate (E1) at (0.7,0.8);
      \coordinate (E2) at (0.2,0.15);
      \coordinate (E3) at (0.4,0.6);
      \coordinate (E4) at (0.4,.95);
      \coordinate (E5) at (0,.95);
      \path [name path=axisy] (axis cs:0,0) -- (axis cs:0,0.95);
      \path [name path=axisx] (axis cs:0,0) -- (axis cs:.95,0);
      \addplot [gray!50!white,name path=A,domain=0:0.45] {2*x+0.05};
      \addplot [gray!50!white,name path=B,domain=0:0.95] {.2*x+0.05};
      \addplot[gray,semitransparent] fill between [of=A and axisy];
      \addplot[gray,semitransparent] fill between [of=B and axisx];
      \draw [thick,latex-latex] (E0) .. controls  (E2) and (E3) .. (E1);
      \fill (E1) circle (1pt);
      \node [right] at (E1) {$\mathrm{E}_1$};
    \end{axis}
  \end{tikzpicture}
  \caption{Cualquier máquina de Carnot reversible operando en $T=0$ debe ser un viaje de ida y vuelta desde un estado de equilibrio dado $\mathrm{E}_1$ a $\mathrm{E}_0$ en $T=0$. Usando la anulación de  $c_x$, la entropía en  $T=0$ se ha dispuesto en un valor finito del eje $S$.}
  \label{fig:nernst}
\end{figure}

En resuemen, $T=0$ sería asignado a un sistema de densidad finita solo si la máquina reversible de Carnot que intenta establecer su temperatura consiste en un viaje de ida y vuelta que conecta un estado  $\mathrm{E}_1$ de la sustancia de trabajo, con un estado $\mathrm{E}_0$, véase la figura~\ref{fig:nernst}.  El ciclo, puramente conceptual, no encierra ningún área, ningún trabajo se produce, y ninguna entropía es intercambiada.\footnote{El ciclo de Carnot se representa habitualmente por un rectángulo en el diagrama $ST$. En realidad, un ciclo de Carnot ---un ciclo que opera con dos repositorios de calor distintos--- requiere de dos procesos isotermos y de dos caminos paralelos que difieren una desplazamiento de entropía $\Lambda$  as shown in Figure~\ref{fig:nernst}.\cite{Martin-Olalla2003b}. En el límite $\Lambda\to0$ los caminos se funden en un viaje de ida y vuelta.  Observe que, conceptualmente, la máquina continua operando con $T_c=0$. Solo cesa de suministrar trabajo.}

La interrelación entre el postulado de Planck del segundo principio y el teorema de Nernst merece también una discusión.

El teorema de Nernst se puede describir por magnitudes relacionadas con la máquina de Carnot reversible. Primero, el límite $T\to0$ es el límite $\mathsf{Q}_c\to0^-$, de la ecuación~(\ref{eq:6}). Segundo, la observación  $(\Delta S)_T\to0$ se puede convertir en $\mathsf{Q}_h\to0^+$ y en $\mathsf{W}\to0^+$. Por tanto el teorema de Nernst establece las siguientes relaciones:
\begin{subequations}
   \label{eq:4}
 \begin{align} 
\mathrm{N}\Longrightarrow& \big((\mathsf{Q}_c\to0^-)\Rightarrow(\mathsf{W}\to0^+) \big),   \label{eq:4a}\\
\neg\mathrm{N}\Longleftarrow&\big( (\mathsf{Q}_c\to0^-)\land \mathsf{W}>0\big).   \label{eq:4b}
 \end{align}  
\end{subequations}

Comparado con la implicación material a derecha de~(\ref{eq:2a}), la implicación material a derecha de~(\ref{eq:4a}) dice: el consecuente ($\mathsf{Q}_c$) se anula solo si el antecedente  ($\mathsf{W}$) se anula también.

Por otra parte, las implicaciones materiales a izquierda de~(\ref{eq:2b}) y de~(\ref{eq:4b}) difieren ligeramente. En (\ref{eq:2b}), $\mathsf{Q}_c$ es una variable categórica: si $\mathsf{Q}_c=0$ y $\mathsf{W}>0$, entonces el postulado se niega; si $\mathsf{Q}_c$ is negativa, entonces el postulado se satisface, sin importar como de cerca a cero pueda estar $\mathsf{Q}_c$. Las refutaciones de Einstein y Epstein usan esta argumentación: es imposible construir una máquina operando en $T=0$ pero, para cualquier $T>0$, sin importar cómo de cerca se esté respecto del cero absoluto, la máquina es posible ya que el argumento categórico que presentaron no se sostiene ya. El postulado de Planck se satisfacería en tanto que $\mathsf{Q}_c<0$.\footnote{Dado que el cambio de entropía isotermo en el refrigerador debe ser negativo, el siguiente argumento categórico también es posible. Para $T_c=0$ el proceso isotermo es también adiabático. Como la entropía no pude disminuir adiabáticamente se deduce que la máquina no podría construirse. Para cualquier temperatura $T_c\neq0$ el argumento ya no vale porque el proceso isotermo es ya diatermo, y la entropía puede de verdad disminuir en este tipo de procesos, independientemente de cómo de cerca la temperatura de la isoterma cero pueda estar $T_c$.}

Por contra, la proposición~(\ref{eq:4b}) usa $\mathsf{Q}_c$ como una variable continua. En esta línea de argumentación, el teorema de Nernst reconoce que el efecto energético $\mathsf{Q}_c$ que debe aparecer en la conversión del calor en trabajo no es categórico sino, como cualquier efecto obligado, tiene que ser una cantidad medible, mayor que un valor prescrito por la naturaleza a través de $\mathsf{Q}_h/T_h$ (la entropía transportada por la áquina) y las propiedades termofísicas de la sustancia que realiza el ciclo.\cite{Martin-Olalla2003b}

%

\appendix

\section{Historia del estudio}
\label{sec:history-manuscript}

La demostración tuvo su origen en la escrituación de un manual de termodinámica que inicié en otoño del año 2023. Durante la vacación de Navidad del año 2023 estaba escribiendo las consecuencias del postulado de Planck, incluyendo el concepto de temperatura. Advertí de la importancia del cero natural de temperatura en relación con el teorema de Nernst e imaginé la demostración inmediatamente.

El estudio fue publicado el 8 de enero del 2024 \doi{10.48550/arXiv.2401.04069} y enviado a revistas punteras.

La demostración se presentó por primera vez el 6 de noviembre del 2024 cuando discutía las consecuencias del postulado de Planck del segundo principio a los alumnos de segundo curso de la asignatura de Termodinámica que habitualmente imparto.
